\documentclass[%
 reprint,
groupedaddress,
showpacs,
 amsmath,amssymb,
 aps,
prb,
]{revtex4-1}

\usepackage{graphicx}
\usepackage{dcolumn}
\usepackage{bm}
\usepackage{hyperref}
\usepackage{amsthm} 
\usepackage[mathlines]{lineno}
\usepackage{color}

\begin{document}

\preprint{APS/123-QED}

\title{Mean-Field Phase Diagram of the Bose-Fermi Hubbard Model}%

\author{Marin Bukov}
\affiliation{%
 Department of Physics, Boston University, 590 Commonwealth Ave., Boston, MA 02215, USA}

\author{Lode Pollet}

 \affiliation{%
  Department of Physics, Arnold Sommerfeld Center for Theoretical Physics, and Center for Nanoscience,\\ Ludwig-Maximilians-Universit\"at M\"unchen, Theresienstra{\ss}e 37, 80333 Munich, Germany}%

\date{\today}

\begin{abstract}

We analyze the ground state properties of mixtures consisting of scalar bosons and spin-$1/2$ fermions using a mean-field treatment of the local boson-fermion interaction on a simple cubic lattice. In the deep superfluid limit of the boson sector and the BCS regime of the fermion sector, we derive BCS-type equations to determine the phase diagram of the system. We find a competition between a charge density wave and a superconducting phase. In the opposite limit, we study the Mott insulator to superfluid transition of the boson sector in the presence of a staggered density-induced alternating potential provided by the fermions, and determine the mean-field transition line. In the two-superfluids phase of the mixture we restrict to nearest-neighbor induced interactions between the fermions and consider the extended Hubbard model. We perform a mean-field analysis of the critical temperature for the formation of boson-assisted $s$-, extended $s^-$-, $d$-, and $p$-wave pairs at fermionic half filling. We compare our results with a recent dynamical mean-field study [Anders \emph{et al.} Phys. Rev. Lett. {\bf 109}, 206401].

\end{abstract}

\keywords{Suggested keywords}

\pacs{67.85.Pq, 71.10.Fd} 

\maketitle

\section{\label{sec:level1}Introduction} 

Bose-Fermi mixtures are ubiquitous in nature. In particle physics, fermions are the constituents of matter interacting via boson-mediated forces. In conventional superconductors, electrons feel an effective attraction due to the retarded interaction with phonons. In $^3$He-$^4$He mixtures, the magnitude of the interactions between the two isotopes is similar to the one of the inter-isotope interactions, which makes it hard to establish a hierarchy of energy scales. Cold atom systems~\cite{Greiner_Bloch} allow for a remarkable control over the interaction strengths and hopping matrix elements which can be fine-tuned to achieve quantum simulation of all these effects. A profound theoretical treatment of Bose-Fermi mixtures at similar interaction energies and length scales is not feasible at the moment.

Spin-polarized Bose-Fermi mixtures have recently been realized experimentally using cold bosonic $^{87}$Rb and fermionic $^{40}$K atoms in three-dimensional (3D) optical lattices. Pioneering experiments~\cite{exp1} in the presence of a fixed attractive background scattering length found a decrease in the bosonic visibility upon increasing the lattice depth, suggesting a shift in the Mott-insulator-to-superfluid (MI-SF) transition. Possible explanations for this effect include self-trapping~\cite{PhysRevLett.96.180403,PhysRevLett.101.050402,
PhysRevB.80.054511,fermipolaron}, adiabatic heating~\cite{PhysRevLett.96.180402,PhysRevLett.100.140409,PhysRevA.77.023608}, or
corrections due to higher bands~\cite{PhysRevB.80.054511,mering}. Another experimental study found an asymmetry between the strongly repulsive and attractive interspecies interactions by analyzing the $^{87}$Rb momentum distribution function~\cite{exp1}.

Theoretically, thermodynamic properties of the dilute Bose-Fermi mixture have been investigated perturbatively, including calculation of the ground state energy, the bosonic momentum distribution function, and the superfluid and normal fractions. It was proposed that by varying the mass ratios and the scattering lengths of the species, it is possible to suppress the bosonic momentum distribution function for small momenta~\cite{viverit2}. 

The phase diagram of Bose-Fermi mixtures with spin-polarized fermions has been investigated in the case of mass and density imbalance between the species both using diagrammatic methods~\cite{pieri1} and the fixed-node diffusion Monte Carlo method~\cite{pieri2}. A first-order quantum phase transition has been found from a polaronic to a molecular state with increasing the boson-fermion scattering length. In the presence of a fixed attractive interspecies interactions the phase diagram has been studied in the continuum analytically using field-theory methods~\cite{ludwig}. It was shown that a strong boson-fermion attraction allows for the formation of molecules containing atoms of either species. Mixtures consisting of spinless fermions and scalar bosons have also been addressed with quantum Monte Carlo simulations in one dimension, studying the phase diagram at double half filling~\cite{Pollet2005} (later found to be in reasonable agreement with a random-phase-approximation (RPA) study~\cite{Barillier2008}) and the destruction of the Mott insulator at unit bosonic density~\cite{Hebert2007}. The supersymmetric point is exactly solvable and was studied in Ref.~\onlinecite{Imambekov}. 
 
Other works, interested in modeling superconductivity, focused on the boson-mediated interaction induced between the fermions. It was shown that an effective interaction potential can be obtained using the $T$-matrix formalism, and the corresponding $s$-wave scattering length has been calculated with the help of field-theoretical methods \cite{bijlsma}. The frequency dependence of the induced interaction is responsible for retardation effects. However, when the bosonic sound velocity is sufficiently greater than the Fermi velocity, the  frequency dependence can safely be neglected, and the resulting interaction is always attractive. As in conventional superconductors, a weak repulsive fermionic interaction can be overcome in favor of an attractive effective one, leading to a Cooper instability in the $s$-wave channel \cite{viverit2,heiselberg}.

In this paper, we present a self-consistent mean-field (MF) treatment of the spinful Bose-Fermi mixture. The MF decoupling of the boson-fermion density-density term enables us to study the effects of the interspecies interaction on the properties of the bosonic and fermion sectors to first order. We identify parameter regimes where a charge density wave is formed, in the presence of which staggered potential modifications to the fermionic chemical potential can be of similar order of magnitude as superconducting fluctuations.    

We do not restrict our analysis to the $s$-wave channel alone and explore the formation of $s$-, extended $s^-$-, $d$-, and $p$-wave pairs on a simple cubic (s.c.) lattice at finite temperature. We work in the limit of short-ranged boson-induced interactions only in the phase where bosons and fermions are both superfluid, without considering a charge density wave. Interestingly, we find approximate analytical expressions for the $s$- and extended $s^-$-wave critical temperatures in very good agreement with numerical solutions of the mean-field equations. Our approximation is equivalent to studying the extended Hubbard model, which has been analyzed in two dimensions in a similar way in Ref.~\onlinecite{short_micnas} where the gap equations have been solved numerically. 

Before starting with the discussion, let us make a few comments on the two-dimensional study by B{\"u}chler and Blatter~\cite{Buchler2003, Buchler2004}, who likewise were interested in Bose-Fermi mixtures and the possible stability of a supersolid phase. Their fermions were however spin-polarized, and they studied the competition between a density-wave instability (in combination with superfluid bosons) and phase separation driven by poles in the Lindhard response function, and related to van Hove instabilities in the density of states. Here, the phase transitions will have a mean-field nature instead of being driven by instabilities in the density of states. The word supersolid will in this paper be used as follows: it is a phase in which each subsystem is superfluid or superconducting and there is a charge density wave (CDW). The density-wave instability in combination with superfluid bosons will hence not be called a supersolid.

This paper is organized as follows. We first introduce the model in Sec.~\ref{sec:model} and the decoupling of the Bose-Fermi interaction term in Sec.~\ref{sec:decoupling}. In Sec.~\ref{SF_limit}, we analyze the physics of the Bose-Fermi mixture in the deep superfluid limit of the boson sector. The latter is treated with a generalized Bogoliubov approximation to account for the presence of the fermion-induced staggered potential. We study the CDW-SF transition in the fermion sector in the weakly interacting BCS limit using the mean-field equations to determine the phase boundary. In Sec.~\ref{MI_limit}, on the other hand, we assume that the fermions are deep in the BEC limit whose only effect is to create a periodic potential for the bosons. We determine the MI-SF phase boundary in the ($U_{bf},t_b$) plane. Section~\ref{USC} is devoted to a mean-field analysis of unconventional superfluidity in the two-superfluids phase of the mixture where we derive approximate equations for the critical temperature of $s$-, extended $s^-$-, $d$-, and $p$-wave pairing. Finally, we conclude in Sec.~\ref{sec:conclusion} and compare our mean-field results to a recent dynamical mean-field theory (DMFT) study.

\section{\label{sec:model}Model}

The system is described by the Hamiltonian

\begin{eqnarray}
 H =&&\ \ H_{b} + H_{f} + H_{int},\nonumber\\
 H_{b} =&& -t_b\sum_{\langle ij\rangle}b^\dagger_i b_j - \mu_b\sum_i n_i + \frac{U_{bb}}{2}\sum_i n_i(n_i - 1),\nonumber\\
 H_{f} =&& -t_f\sum_{\langle ij\rangle,\sigma}c^\dagger_{i\sigma} c_{j\sigma} - \mu_f\sum_i m_i + U_{ff}\sum_i m_{i\uparrow}m_{i\downarrow},\nonumber\\
 H_{int} =&&\ \  U_{bf}\sum_i n_i m_i.
\label{eq:H}
\end{eqnarray} 

The operator $b^\dagger_i$ ($c^\dagger_{i\sigma}$) creates a boson (fermion of spin $\sigma = \uparrow,\downarrow$) at the lattice site $i$, and the chemical potential $\mu_b$ ($\mu_f$) is used to fix the lattice filling. The hopping matrix element between nearest-neighbor sites for the bosons (fermions) is $t_b$ ($t_f$). The number operator on lattice site $i$ is denoted by $n_i$ for bosons, and $m_i = m_{i\uparrow} + m_{i\downarrow}$ for fermions. The bosons are subject to an onsite repulsion $U_{bb}>0$, while we take attractive on-site interactions for the fermions, $U_{ff}<0$. The density of the bosons will be put to unity. The density of the fermions is fixed at half filling, in which case the sign of the on-site boson-fermion interaction $U_{bf}$ is irrelevant. This can be seen from a particle-hole transformation. Therefore, without loss of generality, we assume a repulsive interspecies interaction, $U_{bf}>0$, throughout the rest of this paper~\cite{footnote1}. As energy unit, we set $t_f = 1$ unless otherwise written. 

At these densities, the ground-state phase diagrams of the pure bosonic and pure fermionic models are as follows. The bosonic model undergoes a quantum phase transition from a superfluid to a Mott insulator at a critical value of the ratio $U_{bb}/t_b = 5.8z$ in mean-field theory~\cite{Greiner_Bloch,stoof}, where $z = 2d$ is the lattice coordination number. The superfluid extends also to finite temperature, unlike the Mott insulator. At higher temperature a normal liquid phase is found. The Fermi-Hubbard model at zero temperature is always in a molecular charge density wave (CDW) phase with equally strong superconducting fluctuations due to the SU(2) pseudospin symmetry~\cite{zhang}. One can distinguish between the BCS limit for weak interactions, and the BEC limit for strong negative interactions, $6t_f/|U_{ff}|\ll 1$~\cite{micnas, esslinger,tamaki}. This phase can exist also at finite temperature, but when temperature is sufficiently increased it undergoes a phase transition to a Fermi liquid~\cite{ketterle}. In the absence of bosons, the sign of $U_{ff}$ is irrelevant at half filling due to the Lieb-Mattis transformation. However, when the fermions are coupled to the bosons, the SU(2) pseudo-spin symmetry is broken to U(1), and the sign of $U_{ff}$ becomes relevant. 

\section{\label{sec:decoupling}Hartree decoupling of the Bose-Fermi interaction}

In a recent DMFT study~\cite{anders}, the calculation was set up in such a way that the global symmetry could be broken spontaneously to the following phases, whose order parameters are given in parentheses: \emph{SF$_\text{b+f}$} [both species superfluid: $\langle b_{k=0} \rangle \sim \mathcal{O}(N) $, $ \sum_i \langle c^\dagger_{i\uparrow}c^\dagger_{i\downarrow}\rangle \neq 0$], \emph{CDW+SF$_{\text{b}}$} [charge density wave and superfluid bosons: $\langle b_{k=0} \rangle \sim \mathcal{O}(N) $, $\langle b_{k=\pi} \rangle \sim \mathcal{O}(N) $, $\sum_{i} (-1)^i\langle m_i \rangle \neq 0$], \emph{CDW+SF$_{\text{f}}$} [charge density wave and superfluid fermions: $\sum_{i} (-1)^i\langle n_i \rangle \neq 0$, $\sum_{i} (-1)^i\langle m_i \rangle \neq 0$, $\sum_i \langle c^\dagger_{i\uparrow}c^\dagger_{i\downarrow}\rangle \neq 0$], and \emph{SS} [supersolid - CDW and both species superfluid: $\langle b_{k=0} \rangle \sim \mathcal{O}(N) $, $\langle b_{k=\pi} \rangle \sim \mathcal{O}(N) $, $\sum_{i} (-1)^i\langle m_i \rangle \neq 0$, $\sum_i \langle c^\dagger_{i\uparrow}c^\dagger_{i\downarrow}\rangle \neq 0$]. Here, $N = \sum_i n_i$ is the total number of bosons~\cite{footnote2}.

It is unknown how severe the local approximation in the DMFT is. Also, the role of the retardation is not  clear. For bosons it was found not to matter at all~\cite{anders,anders3}, except in the close vicinity of a bosonic superfluid-to-Mott-insulator phase transition. In this work, we investigate  what aspects of the phase diagram can be recovered by a static and quadratic mean-field theory, which does not require solving an impurity problem as in DMFT.

We therefore perform a mean-field decoupling in the boson-fermion interaction which allows us to keep track of a possible CDW present in the ground state:
\begin{eqnarray}
 n_i m_i \approx&&\ \  n_i\langle m_i\rangle + \langle n_i\rangle m_i  - \langle n_i\rangle\langle m_i\rangle\nonumber\\
 \langle m_i\rangle =&&\ \ m - (-1)^i\alpha \nonumber\\
 \langle n_i\rangle =&& \ \ n + (-1)^i\eta,
 \label{MF_decoupling}
 \end{eqnarray}
where the numbers $n$ ($m$) give the lattice filling of the bosons (fermions), and $\eta$ ($\alpha$) denote the bosonic (fermionic) CDW amplitude. In doing so, we implicitly assume that the fluctuations in the expectation values of the two number operators $n_i$ and $m_i$ are small. Notice the different sign of the amplitudes due to the repulsive interspecies interaction assumed. Also note that DMFT can treat local but nonquadratic terms such as $n_i^2$ and $n_i m_i$ exactly.

Applying the back-action mechanism~\cite{mering} to obtain second- and higher-order corrections due to the interspecies interaction becomes computationally involved in three dimensions. Therefore, our MF analysis in its present form is incapable of self-consistently tracking down phases including nonvanishing density-density correlators, such as $\langle n_i m_{i\sigma}\rangle$, which are predicted to influence the large-$U_{bf}$ physics~\cite{ludwig}. 

In the present approximation, the two sectors of the BFM couple to one another only via the corresponding density-induced alternating mean-field potentials:

\begin{eqnarray}
 H =&&\ \ H_{b}' + H_{f}' + H_{int}',\nonumber\\
 H_{b}' =&&\ \ H_{b} + U_{bf}\sum_i\left(m - (-1)^i\alpha\right)n_i,\nonumber\\
 H_{f}' =&&\ \ H_{f} + U_{bf}\sum_i\left(n + (-1)^i\eta\right)m_i,\nonumber\\
 H_{int}' =&& -U_{bf}\underbrace{\sum_i \left(n + (-1)^i\eta\right)\left(m - (-1)^i\alpha\right)}_{=\ L^3(mn - \eta\alpha)},
\label{eq:H_decoupling}
\end{eqnarray}
where $L$ denotes the system size. The coupling between the bosonic and the fermion sectors is realized by the dependence of the two CDW amplitudes on one another. We may, therefore, choose $\eta(\alpha)$ as a function of $\alpha$, to be determined by the solution of the mean-field equations. The two subsystems can now be diagonalized separately. Finally, we iterate the numerical solution of the gap equations until convergence is reached.

\section{The BCS limit}
\label{SF_limit}

Applying the mean-field approximation~(\ref{eq:H_decoupling}), it is convenient to first consider the regime where the bosons are deep in the superfluid phase, and boson-boson interactions can be treated in the Bogoliubov approximation (cf.~Appendix~\ref{App1}). Any phase transition driven by the fermions takes places in a superfluid background provided by the bosons. The fermions themselves are assumed to be interacting such that BCS-theory applies~\cite{ketterle}. More precisely, for a s.c.~lattice, the parameter regime is $|U_{ff}|/6t_f\ll 1$~\cite{tamaki}. 

Due to the instantaneous density-density boson-fermion interaction, if by some means charge order is attained, it will appear in both sectors. Therefore, keeping in mind that the bosons merely provide the background, we define three pairing order parameters:

\begin{eqnarray}
\alpha =&& \ \ \frac{1}{L^3}\sum_{k\in\text{BZ},\sigma}  \langle c^\dagger_{k+\pi,\sigma}c_{k,\sigma}\rangle   = \frac{1}{L^3}\sum_{i,\sigma} (-1)^i \langle c^\dagger_{i,\sigma}c_{i,\sigma}\rangle  \nonumber\\
\Delta_{0} =&& \ \ \frac{U_{ff}}{L^3}\sum_{k\in\text{BZ}}  \langle c^\dagger_{k,\uparrow}c^\dagger_{-k,\downarrow}\rangle = \frac{U_{ff}}{L^3}\sum_{i}  \langle c^\dagger_{i\uparrow}c^\dagger_{i\downarrow}\rangle, \nonumber\\
\Delta_{\pi} =&& \ \ \frac{U_{ff}}{L^3}\sum_{k\in\text{BZ}}  \langle c^\dagger_{k+\pi,\uparrow}c^\dagger_{-k,\downarrow}\rangle = \frac{U_{ff}}{L^3}\sum_{i} (-1)^i \langle c^\dagger_{i\uparrow}c^\dagger_{i\downarrow}\rangle. \nonumber \\
\label{eq:ferm_MFgaps}
\end{eqnarray}

The $s$-wave gap function $\Delta_0$ detects the appearance of Cooper pairs. The amplitude of the fermionic CDW $\alpha$ measures the strength of the periodic modulation of the fermionic density. It is related to the CDW gap by $\alpha = 2\Delta_c/|U_{ff}|$. The nonuniformity of the $s$-wave gap function is measured by the order parameter $\Delta_\pi$. It can be shown that $\Delta_\pi$ vanishes identically in the case of half-filling due to particle-hole symmetry~\cite{iskin}. Away from half-filling, however, it is in general non-zero and the state of the fermion sector is usually referred to as a staggered superconductor~\cite{micnas} ($\Delta_{\pi}\neq 0$). The supersolid found at half filling in Ref.~\onlinecite{anders} is characterized by $\alpha\neq 0$, $\Delta_0\neq 0$, and $\Delta_\pi = 0$.

Treating the bosons in the extended Bogoliubov approximation results in a quadratic theory. Similarly, a generalized MF theory for the Fermi-Hubbard model in an alternating potential can be derived along the lines of Refs.~\onlinecite{ostlund,ostlund2}. The resulting quadratic theories are then solved in momentum space (cf.~Appendices~\ref{App1} and \ref{App2}). The thermodynamic potential of the Bose-Fermi mixture (BFM) in this approximation is given by

\begin{widetext}
\begin{eqnarray}
\Omega =&&\ \ \Omega_b + \Omega_f + \Omega_{{\rm int}}, \nonumber\\
\frac{\Omega_b}{L^3}  =&& \ \ n_0\left(-zt_b\cos 2\theta -\mu_b + mU_{bf}  - \alpha U_{bf}\sin 2\theta \right) + \frac{U_{bb}}{2}n_0^2(1 + 4\sin^2 2\theta) \nonumber\\
 &&\ \ +\frac{1}{2L^3}\sum_{k\in\text{BZ$'$,s}}\left[E^{(b,s)}_k - \mu_b + U_{bf} + 2n_0U_{bb}  - \frac{1}{\beta}\log\frac{1 + \coth\left(\frac{\beta}{2} E^{(b,s)}_k\right) }{2}  \right],\nonumber\\
\frac{\Omega_f}{L^3}  =&& \ \ \frac{1}{|U_{ff}|}\left( \left( \frac{|U_{ff}|}{2}\alpha\right)^2 + |\Delta_0|^2 + |\Delta_\pi|^2 \right) - \frac{1}{L^3}\sum_{k\in\text{BZ}',s}\left[E^{(f,s)}_k + (\mu_f - nU_{bf})   + \frac{2}{\beta} \log\frac{1 + \tanh\left(\frac{\beta}{2} E^{(f,s)}_k\right) }{2}  \right] \nonumber\\
\frac{\Omega_{\rm int}}{L^3}  =&& \ \ - U_{bf}\left(mn - \alpha\eta(\alpha)\right),
\label{eq:TDpot}
\end{eqnarray}
\end{widetext}
where $\theta$ determines the population fraction of the $\bm k = 0$ and the $\bm k = \bm{\pi}$ condensates (cf.~Appendix~\ref{App1}). The reduced Brillouin zone is defined by $\text{BZ}' = \{\bm k\in \text{BZ}:\sum_i\cos(k_i)\geq0\}$. The presence of the CDW reduces the translational symmetry of the system, which leads to a larger unit cell and a reduced first Brillouin zone, comprising two bands denoted by the index $s = 1,2$. Their dispersion relations $E^{(b,s)}_k, E^{(f,s)}_k$ are given in Eqs.~(\ref{eq:bos_bands}) and~(\ref{eq:fermi_bands}), respectively.

To determine the phase boundary, we apply the following procedure. First, using the Bogoliubov approximation we solve the Bose-Hubbard model in a staggered field whose strength is set by a parameter $\alpha$. We can extract the dependence of the induced bosonic CDW $\eta(\alpha)$ and the total condensate fraction $n_0(\alpha)$ on the staggering field strength $\alpha U_{bf}$ from the thermodynamic potential $\Omega_b$. Recalling that $(\mu_b,n)$ and $(\eta, \alpha U_{bf})$ are conjugate variables, we have $n = -1/L^3\partial_{\mu_b}\Omega_b$ and $\eta(\alpha) = -1/L^3\partial_{\alpha U_{bf}}\Omega_b$ which yields:

\begin{eqnarray}
 n =&&\ \  n_0(\alpha) - \frac{1}{2L^3}\sum_{k\in\text{BZ$'$,s}}\left[1 - \frac{\partial E^{(b,s)}_k}{\partial\mu_b}\coth\left(\frac{\beta}{2}E^{(b,s)}_k\right) \right]\nonumber\\
 \eta(\alpha) =&&\ \ n_0\sin(2\theta) - \frac{1}{2L^3}\sum_{k\in\text{BZ$'$,s}}\frac{\partial E^{(b,s)}_k}{\partial(\alpha U_{bf})}\coth\left(\frac{\beta}{2}E^{(b,s)}_k\right). \nonumber\\
 \label{boson_eqs} 
\end{eqnarray}
At a fixed bosonic density $n$, the first equation is a self-consistency equation for the total Bose condensate fraction $n_0(\alpha)$ (the chemical potential $\mu_b$ being already fixed, c.f.~App.~\ref{App1}), while the second equation determines the CDW amplitude $\eta(\alpha)$.  Solving these equations is equivalent to integrating out the boson sector. 

Solving the Fermi-Hubbard model in the presence of a staggered field of magnitude $\eta(\alpha)$ due to the bosons [cf.~Eq.~(\ref{eq:H_decoupling})], we derive self-consistent MF gap equations for $\Delta_0$, $\Delta_\pi$, and $\alpha$: 

\begin{eqnarray}
m =&&\ \ \frac{1}{2L^3} \sum_{k\in\text{BZ}',s}\left[ 1+ \frac{\partial E^{(f,s)}}{\partial\mu_f}\tanh\left(\frac{\beta}{2}E^{(f,s)}_k\right)\right]
\label{eq:fermion_number}\\
|\Delta_0|  =&&\ \ -\frac{U_{ff}}{2L^3}\sum_{k\in\text{BZ}',s} \frac{\partial E^{(f,s)}}{\partial|\Delta_0|}\tanh\left(\frac{\beta}{2}E^{(f,s)}_k\right),\label{eq:fermion_delta0}\\
|\Delta_\pi| =&&\ \ -\frac{U_{ff}}{2L^3}\sum_{k\in\text{BZ}',s} \frac{\partial E^{(f,s)}}{\partial|\Delta_\pi|}\tanh\left(\frac{\beta}{2}E^{(f,s)}_k\right),\label{eq:fermion_pi} \\
\alpha =&&\ \ \frac{1}{L^3}\sum_{k\in\text{BZ}',s}\left(\frac{\partial E^{(f,s)}}{\partial\alpha}\right)_{\eta(\alpha)}\tanh\left(\frac{\beta}{2}E^{(f,s)}_k\right).
\label{eq:fermion_alpha}
\end{eqnarray}
The signs of the gap functions in the above equations are valid for filling $m<1$. The case $m>1$ can be dealt with using particle-hole symmetry (c.f.~Appendix~\ref{App2}). The value of the fermionic chemical potential is determined by the number equation, Eq.~(\ref{eq:fermion_number}). At half filling ($m=1$), it reads as $\mu_f = U_{bf}n$. 

Finally, we calculate numerically the free energy of the entire mixture for the available candidate states, and the lowest one determines the ground state for a given set of the model parameters.  

It follows directly from the system of MF gap equations above that a supersolid phase is not possible at half-filling within the current approximation. This is most easily seen in the $T=0$ case, where Eqs.~(\ref{eq:fermion_delta0}) and (\ref{eq:fermion_alpha}) assume the form 
\begin{eqnarray}
\alpha =&&\ \ \frac{2}{N_s}\sum_{k\in\text{BZ}'}\frac{-\frac{U_{ff}\alpha}{2} + \eta(\alpha)U_{bf}}{\sqrt{(\varepsilon^{\text{f}}_k)^2 +  |\Delta_0|^2 + \left(-\frac{U_{ff}\alpha}{2} + \eta(\alpha)U_{bf} \right)^2} },\nonumber \\ \\
 |\Delta_0| =&&\ -\frac{U_{ff}}{N_s}\sum_{k\in\text{BZ}'}\frac{|\Delta_0|}{\sqrt{(\varepsilon^{\text{f}}_k)^2 +  |\Delta_0|^2 + \left( -\frac{U_{ff}\alpha}{2} + \eta(\alpha)U_{bf} \right)^2} }.\nonumber\\
\label{bfm_T=0}
\end{eqnarray}
The above system has only two nontrivial solutions, aside from the normal phase ($\alpha = 0$, $\Delta_0 = 0$). This can be shown in the following way: suppose that we are in a phase where $\Delta_0\neq 0$. We can use the second equation and plug it in the first one to obtain $\alpha = -\frac{2}{U_{ff}}\left(-\frac{U_{ff}}{2}\alpha + \eta(\alpha)U_{bf}\right)$. Simplifying, we are left with $\eta(\alpha) = 0$, whose only solution is $\alpha = 0$, since $\eta(\alpha)$ is a monotonically increasing function~\cite{thesis}. The other solution is found at $\Delta_0 = 0$. Hence, the resulting phase can either be a SF or a CDW, but not a superposition. Notice that the same argument holds for $T>0$ as well.

\begin{figure}[b]
\includegraphics[scale=0.4]{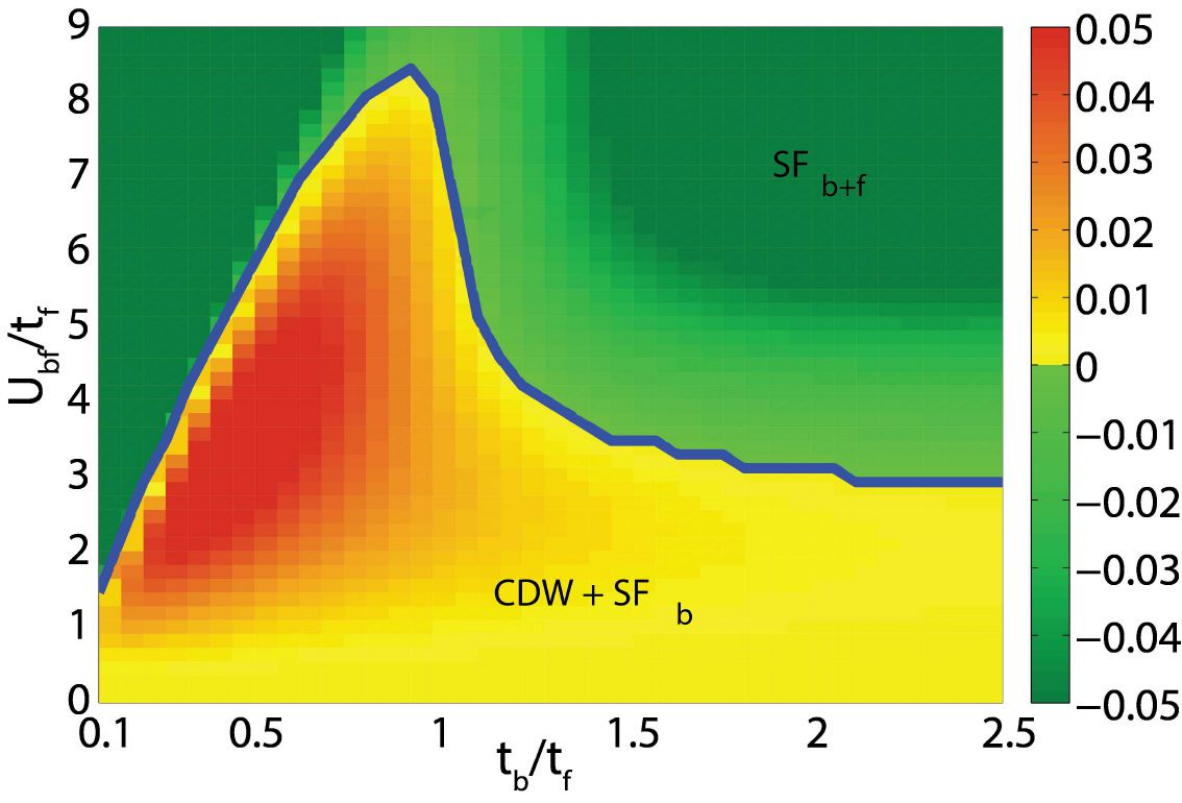}
\caption{\label{fig:MF1}(Color online). Zero-temperature phase diagram of the BFM in the ($U_{bf}$,$t_b$) plane in the presence of a superfluid bosonic background. We see a first order transition where the CDW is lost and the fermions pair into a superfluid. The color bar gives the difference in the grand energies $\Omega$(SF$_{\text{b+f}}$) $-$$\Omega$(CDW+SF$_{\text{b}}$). The model parameters read as $n = 1$, $m = 1$, $T/t_f = 0$, $U_{ff}/t_f = -4$, and $U_{bb}/t_f = 2$.}
\end{figure}

\begin{figure}[b]
\includegraphics[scale=0.45]{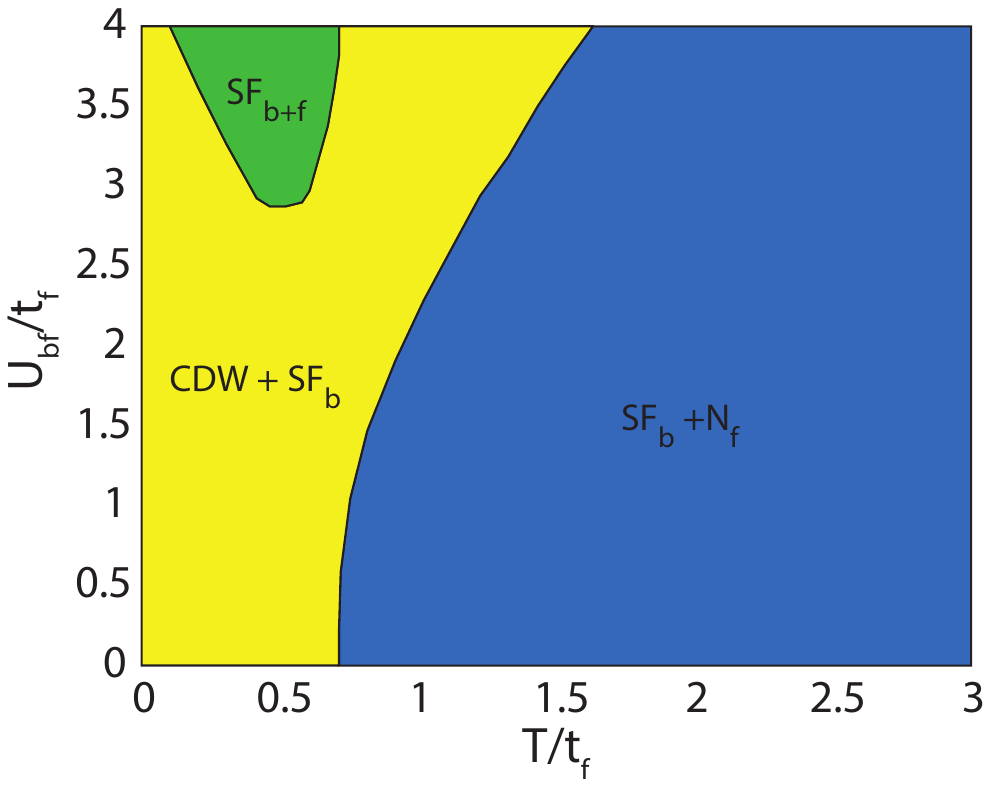}
\caption{\label{fig:MF2}(Color online). Finite-temperature phase diagram of the BFM valid when bosons condense at a much higher temperature at which no signs of fermionic ordering can be seen. For small temperatures the fermion sector can be found in either a SF or a CDW state, depending on the value of $U_{bf}/t_f$. With increasing temperature the system is found either in a CDW+SF$_{\text{b}}$ or in a SF$_b$+N$_\text{f}$ state where the fermions constitute a Fermi liquid, while the bosons are still superfluid. Beyond a critical curve, the CDW order vanishes completely and only the N$_\text{f}$+SF$_\text{b}$ state remains. The parameters are $U_{bb}/t_f = 2$, $t_b/t_f = 1.25$, and $U_{ff}/t_f = -4$.}
\end{figure}

The phase diagram at half filling ($\Delta_\pi = 0$) and $T=0$ is given in Fig.~\ref{fig:MF1}. Starting from low values of $U_{bf}$, the system prefers to be in a state where the bosons have the CDW+SF$_{\text{b}}$ order while the fermions display only the insulating CDW order. Increasing the boson-fermion interaction leads to an at first sight paradoxical phase transition, in which the CDW order is completely lost while the fermions become superfluid. This interplay can be explained by the observation that the boson sector has a higher energy when the CDW is present due to the nonuniformity of the many-body wave function. On the other hand, the fermions can lower their energy when their CDW amplitude increases, since the CDW gap $\Delta_c\sim\alpha$ becomes larger. Taking into account this competition results in a phase transition from a CDW+SF$_{\text{b}}$ state to a SF$_{\text{b+f}}$ state at intermediate values of $U_{bf}$.

In the large $t_b/t_f$ limit, where the bosonic CDW amplitude is negligibly small, the boson and fermion sectors effectively decouple from one another. We expect this to happen when $zt_b\sim\alpha U_{bf}/(1 + n_0 U_{bb}/(\alpha U_{bf}))$~\cite{thesis}, with $z$ the coordination number. Furthermore, a comparison with DMFT~\cite{anders} shows that at small values of $t_b/t_f$ one needs to take into account higher order corrections to the density-density decoupling, such as the boson-mediated attraction. They could influence the functional behavior of the free energy on the order parameters and ultimately alter the slope of the phase boundary. 

Away from half filling, the insulating CDW state for the fermions is no longer found because the Fermi surface no longer exhibits perfect nesting. This has recently been investigated for a fermionic system in Ref.~\onlinecite{iskin}, where a staggered superconducting state (characterized by an alternating non-vanishing gap function, i.e.~$\Delta_\pi\neq 0$) was induced by applying an external alternating potential. Based on this analysis, it follows that a transition occurs between the staggered superconductor and a pure SF state at high values of $U_{bf}$. This is in agreement with earlier work on the extended Hubbard model~\cite{micnas} where the staggered superconductor is stabilized by repulsive nearest-neighbor interactions.

Finite but small temperatures (see Fig.~\ref{fig:MF2}) do not change the physics of the transition, although they modify the values of the order parameters, and hence the position of the transition line. It follows from the gap Equations~(\ref{eq:fermion_alpha}) and~(\ref{eq:fermion_delta0}) that $\Delta_0\leq\Delta_c$ due to the presence of the CDW. The difference in the two gaps is due to the boson-induced potential, and enters the equation as a negative additive correction $-U_{bf}\frac{\mathrm{d}\eta}{\mathrm{d}\alpha}\Big|_{\alpha = 0}$ to the fermion-fermion interaction~\cite{thesis}. Taking this and the exponential dependence of the critical temperature on the inverse interaction strength in BCS theory~\cite{ketterle} into account,  we have at half filling
\begin{equation}
\frac{T_c^\text{CDW}}{T_c^\text{SF}} = \frac{ \exp\left(-\frac{1}{\mathcal{N}_0(|U_{ff}| +  U_{bf}\eta'(0) )} \right) }{\exp\left(-\frac{1}{\mathcal{N}_0(|U_{ff}| )}\right) }\geq 1,
\end{equation}
with equality only for $U_{bf} = 0$. Here $\mathcal{N}_0$ denotes the density of states at the Fermi surface. Physically, this means that a larger gap requires a higher temperature to close. Hence, it follows that the SF order in the fermion sector vanishes first at some $T_c^{(1)}$ when increasing the temperature. Therefore, for $T\geq T_c^{(1)}$ the fermions can either stay in an insulating CDW state for large values of $U_{bf}/t_f$, in which case the BFM is found in the CDW+SF$_{\text{b}}$ state, or make a transition to a Fermi liquid for small values of $U_{bf}/t_f$ so the mixture is found in the SF$_\text{b}$+N$_\text{f}$ state. It is the boson sector which determines the phase of the fermions: a strong CDW is energetically penalized by the delocalized bosons at high $U_{bf}$ as long as temperature allows for a non-zero gap $\Delta_0$. Beyond a certain critical line, the CDW gap also closes, meaning that the fermions are found in the normal Fermi liquid phase N$_\text{f}$ only, while the bosons are still superfluid. Finally, another critical temperature $T_c^{(2)}$ determines the SF-normal transition of the boson sector, above which the BFM leaves the quantum degeneracy regime (not shown). 

In this treatment, we tacitly assumed that a transition to a SF or a CDW in the fermion sector is not destroying the SF order of the bosons. Indeed, a coexistence of two superfluids is stable due to the uniformity of the average density profiles. A strong CDW, on the other hand, may cause a localization of the bosons on a single sublattice, thereby destroying the SF background. However, we find that a CDW state with large density modulations is penalized at large $U_{bf}/t_f$, as it inevitably increases the kinetic energy of the total system compared to the two-SF state. 

\section{The BEC limit}
\label{MI_limit}

In this section, we turn our attention to the regime where the bosons are undergoing a superfluid-to-Mott-insulator transition. A recent DMFT study~\cite{anders} suggests that the fermion sector provides a CDW background in this parameter regime. To model this, we assume that the fermions are deep in the BEC limit ($|U_{ff}|/6t_f\gg 1$) and form locally bound tight molecules sitting on alternating sites, which effectively creates a staggered potential term. We set $\alpha = 1$ for simplicity. This, along with the possibility that the fermion sector be in the superfluid state, may affect the precise position of the transition line. However, we show here that the transition line is driven by the bosons alone. Our aim is to derive a generalization to the MI-SF phase boundary of the boson sector in the presence of the fermion-induced staggered potential. 

By using a cumulant expansion in the bosonic hopping parameter, the phase boundary can be obtained perturbatively from field-theoretical methods~\cite{bradlyn,dossantos}. We write the Hamiltonian as $H_{b}' = H_0 + H_1$ where the perturbation $H_1 = -t_b\sum_{\langle ij\rangle}b^\dagger_i b_j$, and $H_0$ is local and diagonal in the Fock basis. Ignoring retardation effects, it suffices to determine the zero-frequency Green's function of the unperturbed system $G^{(0)}_{A/B}(i\omega = 0) = G^{(0)}_{A/B}(0)$ on each of the sublattices $A$ and $B$:

\begin{eqnarray}
 G^{(0)}_A(0) =&&\ \ \frac{1}{\mathcal{Z}^0_A}\sum_{n=0}^\infty e^{-\beta f_A(n)}\times\nonumber\\
  && \ \ \Bigg[ \frac{2U_{bf}-U_{bb} - \mu_b}{(2U_{bf}-\frac{U_{bb}}{2} - \mu_b +U_{bb}n)^2 - \left(\frac{U_{bb}}{2}\right)^2}\Bigg],  \nonumber\\
 G^{(0)}_B(0) =&& \ \ \frac{1}{\mathcal{Z}^0_B}\sum_{n=0}^\infty e^{-\beta f_B(n)}\times\nonumber\\
 && \ \ \Bigg[\frac{-\mu_b-U_{bb}}{(-\mu_b+U_{bb}n)(-\mu_b-U_{bb} + U_{bb}n)}\Bigg],
\label{eq:bos_greens}
\end{eqnarray}
where $\mathcal{Z}^0_{A/B}$ is the partition function with respect to the local Hamiltonian $H_0$ on sublattices $A$ and $B$, and the grand energies $f_{A/B}$ are given by $f_A(n) = (2U_{bf}-\mu_b)n + \frac{U_{bb}}{2}n(n-1)$ and $f_B(n) = -\mu_b n + \frac{U_{bb}}{2}n(n-1)$.

The finite-temperature effective potential ignoring retardation effects ($\omega = 0$) takes the form~\cite{bradlyn}
\begin{equation}
 \Gamma[\psi] = \frac{L^3\psi^2}{2}\left(\frac{1}{G_A^{(0)}(0)} + \frac{1}{G_B^{(0)}(0)}\right) - \psi^2zt_bL^3 + O(\psi^4).                                     \label{eq:bos_effpot}  
\end{equation}
For a detailed discussion on the Bose-Hubbard related models at finite temperature, see Ref.~\onlinecite{bradlyn}.

At this place we assumed that the effect of the staggered density on the order parameter is negligible in the immediate vicinity of the transition line as approached from within the SF phase. In principle, one would need to distinguish between the values the order parameter takes on the sublattices $A$ and $B$. To argue that this is a subdominant effect, notice that there is indeed a coupling of the form $\psi_A^2\psi_B^2$, but it is of fourth order, and hence can be safely neglected for $\psi$ near the phase boundary.

To find the phase boundary for a second-order transition, we need to set the coefficient in front of the quadratic term to zero. At $T = 0$ this yields
\begin{eqnarray}
 2zt_b \stackrel{!}{=}&&\ \ \frac{1}{G_A^{(0)}(0)} + \frac{1}{G_B^{(0)}(0)} \nonumber\\
       =&&\ \ \frac{-2\mu_b^2 + 2\mu_b(2n-1)U_{bb}-(n-3)n U_{bb}^2}{\mu_b + U_{bb}} \nonumber\\
       &&\ \ - \frac{n(n+1)U_{bb}^2}{\mu_b + U_{bb} - 2U_{bf}} + 2U_{bf}.
\end{eqnarray}
For $U_{bf}=0$, the expression reduces to the well-known result~\cite{stoof}. The condition for unit filling is approximately satisfied with the choice of $\mu_b \approx \frac{U_{bb}}{2} + U_{bf}$. Hence,
\begin{equation}
 U_{bf}\big|_{\text{critical}} \approx \frac{U_{bb}}{2}\sqrt{\frac{6zt_b - U_{bb}}{\frac{2}{3}zt_b - U_{bb}}}.
 \label{eq:bos_MI-SF_dary}
\end{equation}

\begin{figure}[b]
\includegraphics[scale=0.6]{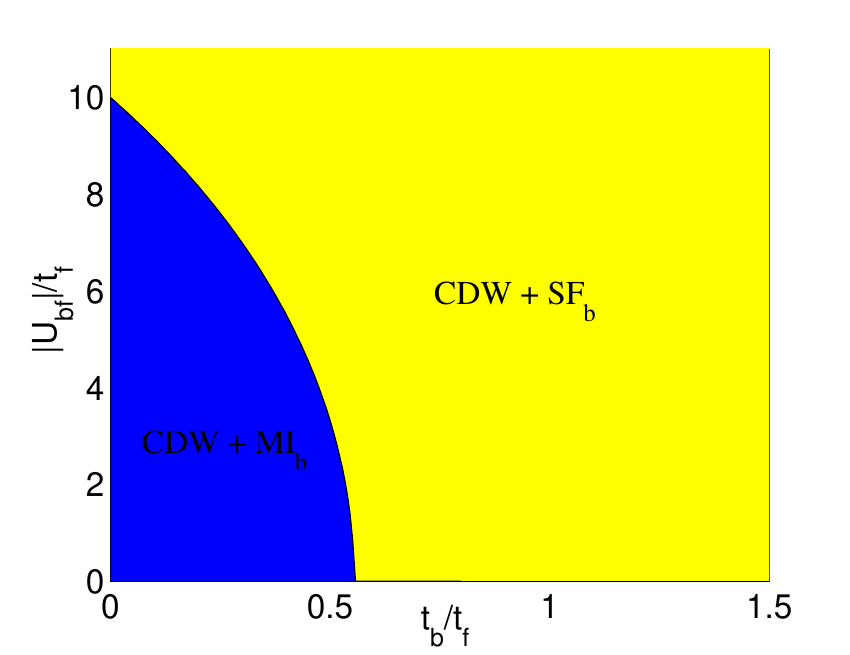}%
\caption{\label{fig:MF3}(Color online). Modification of the bosonic MI-SF transition line in the presence of an alternating potential. The model parameters are $n = 1$, $m = 1$, $U_{bb}/t_f = 20$,  and $T/t_f = 0$.  The alternating chemical potential is generated by freezing locally-paired fermions in a CDW. This order will be picked up by the bosons for all finite interactions and hopping amplitudes.  }
\end{figure}

The phase boundary for $T=0$ is shown in Fig.~\ref{fig:MF3}. 
For $t_b=0$, the value $U_{bb}/2$ is special, since increasing $U_{bf}$ beyond it makes it energetically favorable for the static bosons to form an insulating CDW with a double occupancy on every other site. At any finite $t_b$ the bosons delocalize. On the contrary, for $U_{bf}<U_{bb}/2$, a uniform density minimizes the energy of the system to lowest order. At any small but finite $t_b$ the bosons remain insulating. For $U_{bf}=0$, we recover the MI to SF transition at $U_{bb}/zt_b = 6$. Equation~(\ref{eq:bos_MI-SF_dary}) is in good  agreement with the location of the second order phase boundary found by DMFT~\cite{anders}. Therefore, we have strong arguments that the corresponding phase transition is solely initiated by the bosons, while the (possibly also superfluid) fermions merely provide the staggered potential background.

\section{Unconventional Superfluidity in Bose-Fermi Mixtures}
\label{USC}

In the last section of this work before the conclusions, we investigate the different fermionic unconventional pairing mechanisms that can be induced by superfluid bosons in mean-field theory. 
The main question of interest is under what conditions a $d$-wave superfluid can be stabilized.
A similar analysis but in a different context was carried out in Refs.~\onlinecite{wang, short_micnas}.
We restrict our analysis to half filling for the fermions and unit filling for the bosons and consider low enough temperatures to ensure superfluidity in the boson sector. The bosons can then be treated in the Bogoliubov approximation. In perturbation theory, they generate to lowest order an additional density-density interaction between the fermions which scales as $V_{\text{ind}}\sim-U_{bf}^2/U_{bb}$. Notice that if the following analysis were carried out in a staggered background, one would have had to take into account the boson-induced alternating potential which also goes as $U_{bf}\eta \sim U_{bf}^2/U_{bb}$. This procedure results in an effective fermionic Hamiltonian:

\begin{eqnarray}
 H_{\text{eff}} =&&\  \sum_{k,\sigma}(\varepsilon^{\text{f}}_k-\mu_f)c^{\dagger}_{k\sigma}c_{k\sigma} \\
 +&&\ \frac{1}{2L^3}\sum_{\substack{\sigma\sigma'\\ k_1,k_2,Q}}V^{\text{eff}}_{\sigma\sigma'}(k_1-k_2)c^\dagger_{k_1,\sigma}c^\dagger_{-k_1+Q,\sigma'}c_{-k_2+Q,\sigma'}c_{k_2,\sigma}\nonumber
\end{eqnarray} 
where the fermionic lattice dispersion reads as $\varepsilon^{\text{f}}_k = -2t_f[\cos(k_x)+\cos(k_y)+\cos(k_z)]=\ -t_f\gamma_k$. The effective interaction is given by
\begin{equation}
 V^{\text{eff}}_{\sigma\sigma'}(k,\omega) = U_{\text{ff}}\delta_{\sigma,-\sigma'}+V_{\text{ind}}(k,\omega).
\end{equation}
Here, $V_{\text{ind}}(k,\omega)$ is the induced part of the potential obtained in perturbation theory from the bosonic part of the Hamiltonian~\cite{wang}. In the fast-phonon limit, when the phonon velocity $c_{\text{ph}} = \sqrt{2nU_{\text{bb}}t_{\text{b}}}$ is much larger than the Fermi velocity $\bm{v}_{\text{F}} = \partial\varepsilon^\text{f}_k/\partial\bm{k}|_{k = k_\text{F}}$, the frequency dependence of $V_{\text{ind}}(k,\omega)$, and thus any retardation effects, can safely be neglected. This gives 

\begin{equation}
 V_{\text{ind}}(k) = -\frac{U_{\text{bf}}^2}{U_{\text{bb}}}\frac{1}{1+\xi^2(6-\gamma_k)},
\end{equation} 

with $\xi = \sqrt{t_b/2nU_{\text{bb}}}$ the bosonic healing length. In this limit, the bosons induce a purely attractive potential for the fermions that can lead to pairing even when the latter have initially been free, i.e.~,for $U_{\text{ff}}=0$. Moreover, fermions of the same spin $\sigma$ are now also interacting, allowing for the formation of exotic bound states, such as $p$-wave pairs.

Due to the momentum dependence of the potential, the gap function needed for the investigation of the unconventional pairing mechanisms will also exhibit a nontrivial dependence on the momentum $k$, a feature not present in the simplest BCS theory. Since for a general $k$-dependence no exact analytic results can be obtained, owing to the complexity of the gap equation, we choose to investigate the limit of small bosonic healing length, $\xi\ll 1$. In Ref.~\onlinecite{wang} it was argued that this condition can be realized by controlling the bosonic filling fraction $n$, and experimental realizations for Bose-Fermi mixtures of $^{40}$K-$^{23}$Na and $^{40}$K-$^{87}$Rb atoms have been proposed. We expand the potential as follows:

\begin{eqnarray}
V_{\text{ind}}(k) \approx&&\ -\frac{U_{\text{bf}}^2}{U_{\text{bb}}}\left(\frac{1}{1+6\xi^2}+ \frac{\xi^2}{(1+6\xi^2)^2}\gamma_k\right) \nonumber\\ 
=&&\ -W_0 - W_1\gamma_k,
\label{eq:potapprox}
\end{eqnarray} 
with the short-hand notation $W_0 = U_{\text{bf}}^2/U_{\text{bb}}(1+6\xi^2)^{-1}>0$, and $W_1 = U_{\text{bf}}^2/U_{\text{bb}}\xi^2(1+6\xi^2)^{-2}>0$. It is useful to define the singlet pairing effective potential strength as $V = U_{ff} - W_0$. Hence, in this approximation, the effective fermionic model is equivalent to the extended Hubbard model with nearest-neighbor interactions of strength $W_1$.

The gap equation for a generic $k$-dependent gap function $\Delta_{Q,\sigma\sigma'}(k)=L^{-3}\sum_q V^{\text{eff}}_{\sigma,\sigma'}(k-q)\langle c^{\dagger}_{q,\sigma}c^{\dagger}_{-q+Q,\sigma'}\rangle$ is given by 
\begin{equation}\label{gap}
\Delta_{Q,\sigma\sigma'}(k) = -\frac{1}{2L^3}\sum_q V^{\text{eff}}_{\sigma,\sigma'}(k-q)\frac{\Delta_{Q,\sigma\sigma'}(q)}{E_q}\tanh\left(\frac{\beta E_q}{2}\right),
\end{equation}
where $E_k=\sqrt{(\varepsilon^{\text{f}}_k)^2+\Delta_{Q,\sigma\sigma'}^2(k)}$. The spin indices $\sigma,\sigma'$ are merely a handy notation to keep track of which parts of $V^{\text{eff}}_{\sigma,\sigma'}(k)$ enter the gap equation. Below, we use a separate notation to distinguish between singlet and triplet pairing and, therefore, drop them. We note in passing that the momentum sums in this section are always over the first Brillouin zone $[-\pi,\pi]^3$. The gap functions $\Delta_0$ and $\Delta_\pi$ used in Sec.~\ref{SF_limit} are obtained from $\Delta_{Q,\sigma\sigma'}(k)$ in the special cases of $V^{\text{eff}}_{\sigma\sigma'}(k-q) =U_{ff} \delta_{\sigma,-\sigma'}$ for ${\bm Q} = 0$ and ${\bm Q} = {\bm\pi}$, respectively. 

In the following, we choose ${\bm Q}=0$ and distinguish between \textit{singlet} ($s$-, extended $s^-$- and $d$-wave) and \textit{triplet} ($p$-wave) pairing characterized by order parameters obeying the following symmetries~\cite{annett,raghu}:
\begin{eqnarray}
\text{$s$-wave:}\ \ \Delta(k) =&&\ \ \frac{\Delta^s}{\sqrt{8\pi^3}},\\
\text{$s^-$-wave:}\ \ \Delta(k) =&& \ \ \Delta^{s^{-}}\frac{\gamma_k}{\sqrt{48\pi^3}}, \\
\text{$d_{x^2-y^2}$-wave:}\ \ \Delta(k) =&& \ \  \Delta^{x^2-y^2}\frac{\eta_k}{\sqrt{32\pi^3}}, \\ 
\nonumber\\
\text{$p$-wave:}\ \ \vec d(k) =&& \ \ \frac{\Delta^p}{\sqrt{4\pi^3}}(\sin k_x, \sin k_y, \sin k_z) 
\end{eqnarray}

Notice that the gap function amplitudes $\Delta^s$, $\Delta^{s^{-}}$, $\Delta^{x^2-y^2}$, and $\Delta^p$ are $c$-numbers, while their momentum dependence is separated out. The normalization prefactors assure that the functions on the right-hand-side [$\gamma_k = 2(\cos k_x + \cos k_y + \cos k_z),\eta_k = 2(\cos k_x - \cos k_y)$, and $\sin k_x$] are orthonormal within the first Brillouin zone. The extended $s^-$-wave order parameter still preserves the full rotational symmetry of the gap function, but allows for several nodes on the Fermi surface compared to the conventional $s$-wave one. We only need to consider either the $d_{z^2}$ or the $d_{x^2-y^2}$ wave gap function, since they lead to degenerate transition temperatures~\cite{annett}, while it can be shown~\cite{thesis} that the gaps of the $d_{yz}$, $d_{yx}$, and $d_{xz}$ channels  vanish identically in the approximation for the interaction potential of Eq.~(\ref{eq:potapprox}).

Following the discussion in Ref.~\onlinecite{short_micnas}, we define the function $F_k(\beta)=\frac{\tanh(\beta E_k/2)}{E_k}$, where $\beta = 1/T$. The gap equations can be calculated as
\begin{eqnarray}
\text{$s$-wave:}\ \ 1 =&& -V\varphi_1(\beta),\ \ \ \varphi_1(\beta) = \frac{1}{2L^3}\sum_k F_k(\beta), \label{eq:gaeqs0} \\
\text{$s^-$-wave:}\ \ 1 =&&\ \ \frac{W_1}{6}\varphi_\gamma(\beta), \ \ \varphi_\gamma(\beta) = \frac{1}{2L^3}\sum_k\gamma_k^2F_k(\beta), \\
\text{$d$-wave:}\ \ 1 =&& \ \ \frac{W_1}{3}\varphi_\eta(\beta), \ \ \varphi_\eta(\beta) = \frac{1}{2L^3}\sum_k\eta_k^2F_k(\beta), \\
\text{$p$-wave:}\ \ 1 =&& \ \ 2W_1 \varphi_p(\beta), \ \varphi_p(\beta) = \frac{1}{2L^3}\sum_k\sin^2k_xF_k(\beta). \nonumber \\
\label{eq:gaeqs}
\end{eqnarray}

At the inverse critical temperature $\beta_c$ the corresponding gaps vanish. For $\beta_c t_f\gg 1$, we first evaluate the functions $\varphi_i(\beta_c)$ in the thermodynamic limit (c.f.~Appendix~\ref{App3}), leading to

\begin{eqnarray}
 \varphi_1(\beta_c) \approx&&\ \ \mathcal{N}_0\bigg[\frac{9}{2}\log2\log\left(\frac{12e^C}{\pi}\beta_ct_f\right)+\kappa_1\bigg], \label{eq:UCS_phis0} \\
 \varphi_{\gamma}(\beta_c) \approx&&\ \ \mathcal{N}_0\bigg[ - \frac{3\pi^2\log2}{16}\frac{1}{(\beta_ct_f)^2}+\kappa_{\gamma}\bigg], \\
 \varphi_{\eta}(\beta_c) \approx&&\ \ \mathcal{N}_0\bigg[\frac{\pi^2}{4}\left(\frac{\log2}{4}-1 \right)\frac{1}{(\beta_ct_f)^2} \nonumber \\
 &&\ \ +3(9\log2-1)\log\left(\frac{12e^C}{\pi}\beta_c t_f\right)+\kappa_{\eta}\bigg], \\
 \varphi_p(\beta_c) \approx&&\ \ \frac{\mathcal{N}_0}{4}\bigg[\frac{\pi^2}{12}\frac{1}{(\beta_ct_f)^2} \nonumber \\
 && +(9\log2+1)\log\left(\frac{12e^C}{\pi}\beta_c t_f\right)+\kappa_p\bigg], 
 \label{eq:UCS_phis}
\end{eqnarray} 
where $\mathcal{N}_0 = 1/(2\pi^2t_f)$ and $C\approx 0.557$ is the Euler constant. The non-universal, lattice- and dimension-dependent constants read as $\kappa_1 = -1.90$, $\kappa_\gamma = 5.47$, $\kappa_\eta = -3.78$ and $\kappa_p = -6.96$.

The gap equations for $s$-wave and $s^-$-wave can be solved algebraically: 

\begin{eqnarray}
\frac{T_c^s}{t_f} =&&\ \ \frac{12e^C}{\pi}\exp\left[-\frac{2}{9\log2}\left(-\kappa_1+\frac{1}{\mathcal{N}_0|V|}\right)\right]
\label{eq:BCST},\\
\frac{T^{s^-}_c}{t_f}=&&\ \ \sqrt{\frac{16}{3\pi^2\log2}\left(\kappa_{\gamma}-\frac{6}{\mathcal{N}_0 W_1}\right)}. 
 \label{eq:Tcexts}
\end{eqnarray}

We recover the expected exponential decay in the critical temperature from BCS theory for negative interaction including the numerical prefactors for a simple cubic (s.c.) lattice. The extended $s^-$-wave critical temperature is clearly nonperturbative. Furthermore, it follows from Eq.(~\ref{eq:Tcexts}) that there cannot be any extended $s^-$-wave pairing for $W_1<6/(\kappa_{\gamma}\mathcal{N}_0)$ on a s.c.~lattice. This law qualitatively explains the behavior of the corresponding numerical curve calculated for the extended $s^{-}$-wave critical temperature of the extended Hubbard model in two dimensions~\cite{short_micnas}.

The $d$- and $p$-wave pairing functions, $\varphi_{\eta}$ and $\varphi_p$, however, contain both algebraic and logarithmic terms. Therefore, inverting their gap equations to obtain analytical expressions for the critical temperatures is possible only numerically. For $p$-wave pairing, Ref.~\onlinecite{efremov} predicted exponential decay of the critical temperature for a Bose-Fermi mixture in the continuum using a similar expansion technique for the gap function. Hence, the effect of the lattice structure for $p$-wave pairing can be traced back to the appearance of an algebraic term in $\varphi_p(\beta_c)\propto 1/(\beta_ct_f)^2$, and the value of the corresponding constant $\kappa_p$.

\begin{figure}
\begin{minipage}{\columnwidth}
\includegraphics[scale=0.45]{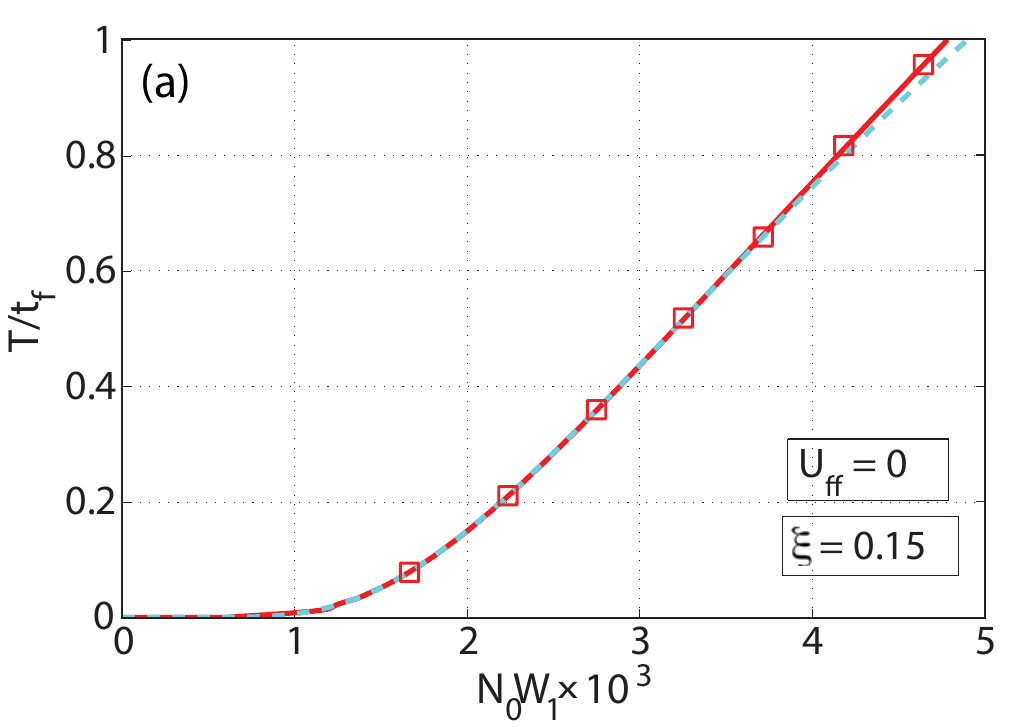}
\end{minipage}
\begin{minipage}{\columnwidth}
\includegraphics[scale=0.45]{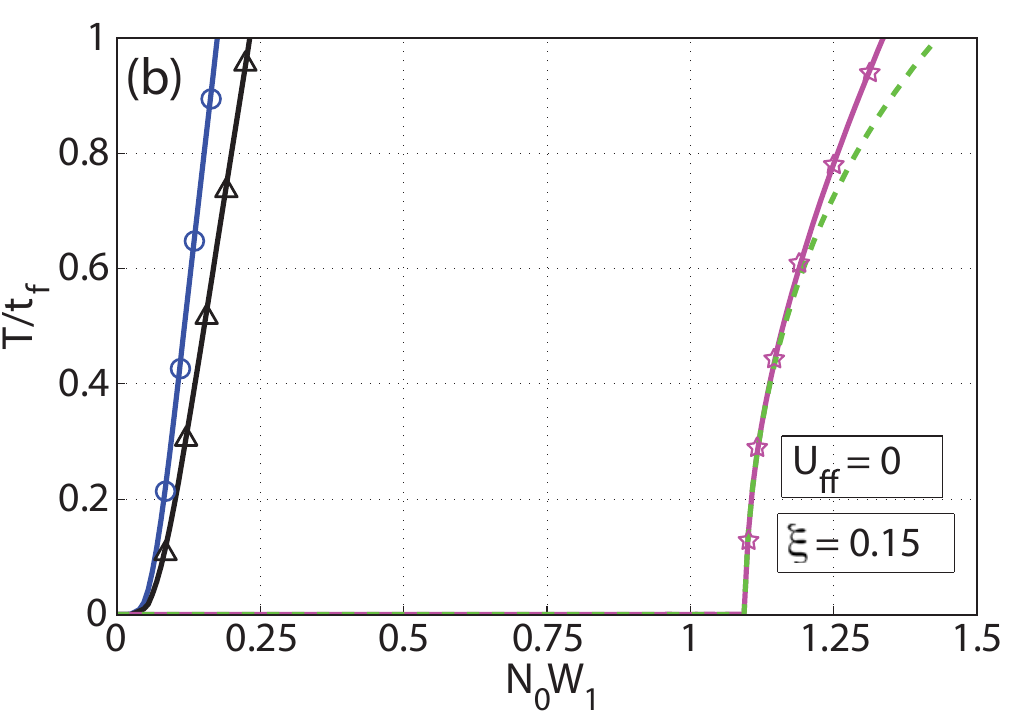}
\end{minipage}
\caption{\label{fig:USC}(Color online) (a) Mean-field critical temperature for $s$-wave pairing (empty squares) as a function of the nearest-neighbor interaction parameter $\mathcal{N}_0W_1 = \mathcal{N}_0\frac{U_{bf}^2}{U_{bb}}\frac{\xi^2}{(1+6\xi^2)^2}$. (b) In the special case of $U_{ff}=0$ one can compare it to the critical temperatures for unconventional pairing: $s^-$- (empty stars), $d$- (empty circle), and $p$-wave (empty triangles). The solid lines represent full numerical solutions of the mean-field equations~(\ref{eq:gaeqs0})-~(\ref{eq:gaeqs}), while the nearby dashed lines are according to our approximate analytical results taking Eqs.~(\ref{eq:potapprox}) and~(\ref{eq:approx_DOS}) into account.}
\end{figure}

A direct comparison between the pairing instabilities for extended $s^-$-, $d$-, and $p$-wave pairing is possible, since their critical temperatures depend on the parameter $W_1$ only. Figure~\ref{fig:USC} (b) shows that $d$- and $p$-wave pairing are possible also for small values of $W_1$, while extended $s^-$-wave pairing is completely suppressed. In Fig.~\ref{fig:USC}, we also compare the quality of the approximations made to the gap equations~(\ref{eq:gaeqs}) with a full numerical evaluation of these equations and find a very good agreement. At very high transition temperatures $T/t_f\sim 1$, the temperature dependence of the gap equation for energies in the upper part of the band (interval $[1,3]$) is no longer negligible (c.f.~Appendix~\ref{App3}). The ratio $t_b/U_{bb} = 0.22$ used is consistent with the assumption that the boson sector is superfluid and this choice was motivated by the parameters used in Fig.~4 of Ref.~\onlinecite{anders}.

The situation becomes more involved if we also take $s$-wave pairing into account, since the $s$-wave interaction $V = U_{ff}-W_0$ contains both a generic ($U_{ff}$) and an induced part ($W_0$). Any direct comparison between all critical temperatures, therefore, depends on a fixed value of $U_{ff}$. For $U_{ff} = 0$, the validity of our approximation requires $W_1\ll W_0$, and hence the $d$-wave pairing is strongly suppressed by the $s$-wave pairing: c.f.~Fig.~\ref{fig:USC} (a). This result agrees well with the observation of Ref.~\onlinecite{wang}, according to which the unconventional pairing mechanisms are mostly pronounced for $\xi\sim 1$. We note in passing that $^{40}$K-$^{23}$Na is an interesting mixture to observe unconventional pairing~\cite{wang}. Figure~\ref{fig:USC} shows the critical temperatures for all the pairing mechanisms in the case $U_{ff} = 0$. The $s$-wave pairing is clearly the dominating one [notice the relative factor between the horizontal axes in Figs.~\ref{fig:USC}(a) and~\ref{fig:USC}(b)], followed by $d$-, $p$- and the $s^-$-wave pairing. The $s^-$-wave pairing is suppressed for small values of $\mathcal{N}_0W_1$, following Eq.~(\ref{eq:Tcexts}). 

An interesting situation shows up when the bare fermionic interaction is positive and approximately equal in magnitude to the induced one: $U_{ff} \approx |W_0|>0$. This could be achieved using Feshbach resonances in systems of ultracold atoms. In this case, the boson-induced attraction compensates exactly for the fermion-fermion repulsion, thereby closing the $s$-wave channel completely, and leaving $d$-wave pairing as the dominant pairing mechanism. This, in turn, opens up the possibility to observe exotic $d$-wave pairs. We remark that dominant $d$-wave pairing has already been predicted in a two-dimensional (2D) Bose-Fermi mixture on an isotropic square lattice using a renormalization group approach~\cite{mathey}.

\section{Conclusion}
\label{sec:conclusion}

We obtained the finite and zero-temperature phase diagrams of the Bose-Fermi-Hubbard model in a mean-field approximation. Treating the boson-fermion interaction along the lines of mean-field theory couples the bosonic and fermion sectors via their CDW amplitudes. One then finds both the Bose- and Fermi-Hubbard models in the presence of a staggered potential, whose strength is determined by the expectation value of the on-site density of the opposite species. This scheme allows us to look for exotic states, characterized by a simultaneous superfluid and crystalline long-range order. 

As a solution to the self-consistency equations of the fermion sector in the limit of superfluid bosons, we recover the familiar SF and CDW phases. We find that a supersolid phase is not possible at half filling in the present MF description. This is related to the difference in the values of the CDW and SF gap induced by the staggered potential. The ultimate ground state of the BFM is determined by comparing the total mean-field free energies of the candidate ground states. We find that MF supports both the CDW+SF$_{\text{b}}$ and the SF$_{\text{b+f}}$ phases for small and large values of $U_{bf}$, respectively. We also successfully calculated a very similar transition line to the one observed in DMFT for the CDW+SF$_{\text{f}}$ to CDW+SF$_{\text{b}}$ transition~\cite{anders}, as a consequence of the bosonic MI-SF transition in the presence of a staggered potential.

In the SF$_{\text{b+f}}$ phase higher order corrections due to the boson-fermion interaction lead to an induced long-range attractive potential between the fermions. Truncating the latter to nearest-neighbors and assuming half filling, we derive the gap equations for $s$-, extended $s^{-}$-, $d$- and $p$-wave pairing. The corresponding critical temperatures are obtained as a special case of vanishing gap parameters. This leads to transcendental equations, except for the cases of $s$- and extended $s^-$-wave pairing where very good analytical approximations to the mean-field gap equations have been developed in the limit $T/t_f\lesssim 1$. 

For $s$-wave, we recover the exponential decay expected from BCS theory with the appropriate prefactor in the mean field-approximation. For extended $s^{-}$-wave, we find a non-perturbative square-root dependence $T_c/t_f\sim\sqrt{1 - 6/(\kappa_\gamma W_1\mathcal{N}_0)}$. Hence, extended $s^{-}$-wave pairing is possible only starting from a critical value $W_1^\text{c} = 6/(\kappa_{\gamma}\mathcal{N}_0)$. Among the extended $s^-$, $d$ and $p$-wave instabilities, the dominant one is the $d$-wave. Taking $s$-wave pairing into account, we identify two interesting scenarios: either one can use the bare fermionic interaction $U_{\text{ff}}$ to effectively sweep the effective $s$-wave interaction parameter $V$ up to zero closing the $s$-wave channel, so that the $s$-wave instability succumbs to the $d$-wave one, or one can consider $U_{\text{ff}}=0$, in which case we recover the observation of Ref.~\onlinecite{wang} that conventional $s$-wave pairing dominates. 

When comparing our results with the DMFT phase diagram of Ref.~\onlinecite{anders} there remain some striking differences, most notably the absence of a supersolid in the present treatment. The corresponding orders were quite robust in the DMFT study, making it unlikely that a numerical artifact has caused it. We also do not think that the retardation is responsible for the difference in phases. Most likely, the crucial difference is in the treatment of the local boson-boson and boson-fermion density interactions, which go beyond the Hartree-type decoupling of the Bose-Fermi interaction term performed here. In DMFT these local correlations are treated exactly, and they are important in localized, commensurate phases such as Mott insulators~\cite{anders, anders3}, while they are completely missing in our approximate quadratic theories.  It may be interesting in future work to address the question as to whether an extended mean-field theory~\cite{deforcrand} or a variational cluster approach~\cite{VCA} can be developed in which local density-density correlations are represented by a variational parameter within one unit cell and if such a theory can reproduce the same shape of the phase diagram as DMFT. Close to half filling, it is also worth noting that both treatments agree and are able to find a supersolid phase.

\begin{acknowledgments}
We wish to thank I. Bloch, H. P. B{\"u}chler and A. Muramatsu for valuable discussions. This work is supported by the Excellence Cluster NIM, FP7/Marie-Curie Grant No. 321918 (”FDIAGMC”), FP7/ERC Starting Grant No. 306897 (”QUSIMGAS”) and by a grant from the Army Research Office with funding from DARPA.
\end{acknowledgments}

\begin{appendix}

\section{The Bose-Hubbard Model in a Staggered Potential: the SF Limit}
\label{App1}

In this appendix, we present the Bogoliubov treatment of the Bose-Hubbard model subject to an alternating chemical potential. For a more-detailed discussion, see Ref.~\onlinecite{thesis}. The Hamiltonian in momentum space is given by
\begin{eqnarray}
 H =&&\ \ \sum_k(\varepsilon^{\text{b}}_k-\mu_b+U_{bf}) b^\dagger_k b_k 
 -\alpha U_{bf}\sum_k b^\dagger_{k+\pi}b_k \nonumber\\  &&\ \ +\frac{U_{bb}}{2L^3}\sum_{k_1,k_2,k_3,k_4}b^\dagger_{k_1}b^\dagger_{k_2}b_{k_3}b_{k_4}\delta_{k_1+k_2,k_3+k_4}. 
 \label{eq:BHM_stagg}
\end{eqnarray}
In the absence of interactions the bosons would undergo a BEC occupying macroscopically the modes $\bm k = 0$ and $\bm k = \bm\pi = (\pi,\pi,\pi) $, since the alternating potential couples directly the modes $\bm k$ and $\bm{k + \pi}$. Thus, in the presence of interactions, one has to modify the Bogoliubov approximation accordingly,
\begin{equation}
 b_k \longrightarrow b_k + \begin{cases}
                             \cos\theta\sqrt N_0\delta_{k,0}\ \ \   &k\in\text{BZ}' \\
                             \sin\theta\sqrt N_0\delta_{k+\pi,0}\ \ \   &k\in\text{BZ$\setminus$BZ}'. \\
                           \end{cases}
\label{eq:bos_bog_bs}
\end{equation}
Here $N_0$ denotes the superfluid fraction, and $\theta = \theta(t_b,\alpha U_{bf}, U_{bb})$ parametrizes the occupation number fraction of the $\bm{k} = 0$ and $\bm k = \bm{\pi}$ condensates: $f_{ k = 0} = n_0\cos^2\theta$, $f_{ k = \pi} = n_0\sin^2\theta$, $n_0 = f_{ k = 0} + f_{ k = \pi}$. The reduced Brillouin zone is defined by $\text{BZ}' = \{\bm k\in \text{BZ}:\varepsilon_k\leq0\}$. To make a clear distinction, we also define $\alpha_k = b_k$, for $\bm k\in \text{BZ}'$, and $\beta_k = b_k$, for $\bm k\in\text{BZ$\setminus$BZ}'$. 

One then follows the standard procedure: plugging Eq.~(\ref{eq:bos_bog_bs}) in Eq.~(\ref{eq:BHM_stagg}) and collecting terms to order $N_0$ yields a Hamiltonian quadratic in the $b_k$ operators. Setting the linear terms in $b_0$ and $b_\pi$ to zero ensures stability and results in a system of equations that fixes $\mu_b$ and $\theta$:

\begin{eqnarray}
(-zt_b - \mu_b + U_{bf})\cos\theta - \alpha U_{bf}\sin\theta&& \nonumber\\
+ n_0 U_{bb}\cos\theta(1+\sin^2\theta&&)=0,\nonumber\\
 (zt_b - \mu_b + U_{bf})\sin\theta - \alpha U_{bf} \cos\theta \nonumber\\
+ n_0 U_{bb}\sin\theta(1+\cos^2\theta&&)=0,
\label{eq:bos_chem_pot}
\end{eqnarray}
where $z = 2d$ is the coordination number. The quadratic part is a generalization of the Bogoliubov Hamiltonian in the absence of a staggered potential. It is useful to introduce the abbreviations $\varepsilon^{\text{b}}_k = -2t_b\sum_i^d\cos(k_i)$, $u = U_{bb}n_0$, $w = U_{bb}n_0\sin 2\theta$, $v =2w - \alpha U_{bf}$, and $\tilde\mu = \mu_b-U_{bf}-2u$. We can now reduce the Brillouin zone defining 
$\bm{\alpha_k}\ = \left(\alpha_k,\beta_k,\alpha^\dagger_{-k},\beta^\dagger_{-k}\right)^t$, to arrive at $H_{\text{Bog}} =\ \text{const} + \frac{1}{2}\sum_{k\in\text{BZ$'$}}\bm\alpha_k^\dagger h_k \bm{\alpha_k}$. The $4\times 4$ matrix $h_k$ is given by 
\begin{equation}
h_k = \left( \begin{array}{cccc}
                                                                                                                       × \varepsilon^{\text{b}}_k-\tilde\mu & v & u & w \\
                                                                                                                         v & -\varepsilon^{\text{b}}_k-\tilde\mu & w & u \\
                                                                                                                         u & w & \varepsilon^{\text{b}}_k-\tilde\mu & v \\
                                                                                                                         w & u & v & -\varepsilon^{\text{b}}_k-\tilde\mu
                                                                                                                      \end{array}\right).
                                                                                                      \nonumber
\end{equation}

This Hamiltonian is canonically diagonalized using a pseudo-unitary transformation~\cite{thesis}, and the two energy bands labeled by $s = 1,2$ are given by

\begin{eqnarray}
 E^{(\text{b},s)}_k = \Bigg(&& v^2 - u^2 - w^2 + \varepsilon_k^2 + \tilde\mu^2  \nonumber\\
 &&\pm 2\sqrt{ \left(v\tilde\mu +  uw\right)^2 + \left(\tilde\mu^2-w^2\right)\varepsilon_k^2  }\   \    \Bigg)^{1/2}.
\label{eq:bos_bands}
\end{eqnarray}
As expected, the lower energy band is linear for $|\bm k|\to 0$, and there is a band gap at $|\bm k| = |\bm\pi|$.

Taking into account the constant terms including those resulting from the commutation relations, it is straightforward to find the thermodynamic potential $\Omega_b$ [cf.~Eq.~\ref{eq:TDpot}].

\section{The Fermi-Hubbard Model in a Staggered Potential: the BCS Limit}
\label{App2}

This appendix deals with the Fermi-Hubbard model in a staggered potential in the BCS-regime. Allowing for CDW and SF order in the ground state, Wick's theorem yields the following mean-field decoupling for the interaction part of the Hamiltonian~\cite{ostlund,ostlund2,thesis}:

\begin{eqnarray}
U_{ff}\sum_i m_{i\uparrow}m_{i\downarrow}\approx&& \ \ \Delta_c\sum_{k,\sigma}  c^\dagger_{k+\pi,\sigma}c_{k,\sigma}\nonumber\\
&& + \sum_k \left(\Delta_0 c_{-k,\downarrow}c_{k,\uparrow} + \text{H.c.}\right)\nonumber\\
&& + \sum_k \left(\Delta_\pi c_{-k,\downarrow}c_{k + \pi,\uparrow} + \text{H.c.}\right) \nonumber\\
&& - \frac{L^3}{U_{ff}}\left( \Delta_c^2 + |\Delta_0|^2 + |\Delta_\pi|^2\right),
\end{eqnarray}
where the CDW and s-wave pairing order parameters are, respectively, defined in Eqs.~(\ref{eq:fermion_delta0})-(\ref{eq:fermion_alpha}), and $\alpha = 2\Delta_c/|U_{ff}|$.

As for bosons, we reduce the Brillouin zone and define the operators $\alpha_k = c_k$, for $\bm k\in \text{BZ}'$, and $\beta_k = c_k$, for $\bm k\in\text{BZ$\setminus$BZ}'$. Defining a Nambu spinor $\bm{\alpha}_k = (\alpha_{k\uparrow},\beta_{k,\uparrow},\alpha^\dagger_{-k\downarrow},\beta^\dagger_{-k\downarrow})^t$ valid in the reduced Brillouin zone, one can write the full mean-field Hamiltonian, including the staggered potential and the kinetic energy, as $H = \text{const} + \sum_{k\in\text{BZ}'}\bm\alpha^\dagger_k h_k \bm\alpha_k$, where
\begin{equation}
h_k = \left(\begin{array}{cccc}
      ×\varepsilon^{\text{f}}_k - \mu & \Delta_c+\eta   U_{bf} & |\Delta_0| & |\Delta_\pi| \\
       \Delta_c +\eta U_{bf} & -\varepsilon^{\text{f}}_k - \mu & |\Delta_\pi|  & |\Delta_0| \\
       |\Delta_0| & |\Delta_\pi|  & -\varepsilon^{\text{f}}_k + \mu & -\Delta_c-\eta U_{bf} \\
       |\Delta_\pi|  & |\Delta_0| & -\Delta_c-\eta U_{bf} & \varepsilon^{\text{f}}_k + \mu
     \end{array}\right),
\end{equation} 
where $\varepsilon^{\text{f}}_k = -2t_f\sum_i^d\cos(k_i)$ and $\mu = \mu_f - U_{bf}n$ ($n$ is the bosonic filling). Here, we used that for fermionic filling $m\leq 1$, $\Delta_0$ and $\Delta_\pi$ can be assumed real and positive, which is not a restriction due to particle-hole symmetry~\cite{iskin}. The band structure reads as
\begin{eqnarray}
E^{(\text{f},s)}_k =&&\ \  \Big( (\Delta_c + \eta U_{bf})^2 + |\Delta_0|^2 + |\Delta_\pi|^2 \nonumber\\
&&\ \ + (\varepsilon^{\text{f}}_k)^2 + \mu^2 \pm 2 X \Big)^{1/2} , \nonumber\\
 X =&&\ \ \Big( \left(|\Delta_0||\Delta_\pi| + \mu_f(\Delta_c + \eta U_{bf}) \right)^2  \nonumber\\
 &&\ \ + (\varepsilon^{\text{f}}_k)^2\left(|\Delta_\pi|^2 + \mu_f^2 \right)  \Big)^{1/2}.
 \label{eq:fermi_bands}
\end{eqnarray}
As in the bosonic case, the grand potential $\Omega_f$ [c.f.~Eq.~\ref{eq:TDpot}] can now be calculated if one includes the constant terms coming from commutator relations.

\section{Calculation of the functions $\varphi_i$}
\label{App3}

In this appendix, we give the details of the calculation of the auxiliary functions $\varphi_i$, which enter the gap equations~(\ref{eq:gaeqs}). The essence of our approximation can be summarized as follows:

\begin{enumerate}
 \item We pass to the thermodynamic (TD) limit.
 \item We make use of a logarithmic approximation to the 2D density of states (DOS) $N_{\text{2D}}(\varepsilon)$ to define the 3D DOS $N_{\text{3D}}(\varepsilon)$ via
 \begin{equation}
 N_{\text{3D}}(2t_f\gamma) = \frac{1}{\pi}\int_{\text{max}\{-2,\gamma-1\}}^{\text{min}\{2,\gamma+1\}}d w\frac{N_{\text{2D}}(2t_fw)}{\sqrt{1-(\gamma-w)^2}}. \nonumber
 \end{equation}
 \item Due to the structure of $N_{\text{3D}}(\varepsilon)$, we use suitable approximations to $F_p(\beta_c)$ to divide the dimensionless half bandwidth interval (to be integrated over) in two pieces: $[0,3] = [0,1]\cup[1,3]$.
 \item We keep the full temperature dependence of $F_p(\beta_c)$ in the interval $[0,1]$ where exact results for the 3D DOS can be obtained, while putting $\beta_ct_f\gg 1$ in $[1,3]$, thus replacing $\tanh(\beta_c\varepsilon/2)\approx 1$. Hence, we ignore the temperature contribution of the complicated part of $N_{\text{3D}}(\varepsilon)$ in the interval $[1,3]$ evaluating the resulting integral to a non-universal constant.
\end{enumerate} 

Let us now be specific: the 2D DOS can be approximated to a very good accuracy by a logarithm:

\begin{equation}
 N_{\text{2D}}(\varepsilon) = \frac{2}{D\pi^2}K\left(\sqrt{1-\left(\frac{\varepsilon}{D}\right)^2}\right)\approx \frac{2}{D\pi^2}\log\left|\frac{4\sqrt2D}{\varepsilon}\right|,
 \label{eq:approx_DOS}
\end{equation} 

with $K(x)$ the complete elliptic integral of the first kind and $D=4t_f$ the half bandwidth. Within the same accuracy, the 3D DOS reads

\begin{displaymath}
   N_{\text{3D}}(\varepsilon) \approx \left\{
     \begin{array}{lr}
       \mathcal{N}_0\frac{9}{2}\log2, &  0\leq\gamma \leq1\\
       \displaystyle{\frac{\mathcal{N}_0}{\pi}\int}_{\gamma-1}^{2}dw\frac{\log\left|\frac{8\sqrt2}{w}\right|}{\sqrt{1-(\gamma-w)^2}}, &  1\leq\gamma\leq3
     \end{array}
     \right.
\end{displaymath}
where $\varepsilon = 2t_f\gamma$, and $\mathcal{N}_0 = 1/(2t_f\pi^2)$.

Now we are ready to proceed towards the calculation of $\varphi_1$. At the critical temperature the gap closes and we have $E_k = \sqrt{\varepsilon_k^2+\Delta_k^2} = \varepsilon_k$. Using this, we compute

\begin{widetext}
\begin{eqnarray}
 \varphi_1(\beta_c) =&&\ \ \frac{1}{2L^3}\sum_k F_k(\beta_c) \stackrel{\text{TD-limit}}{\longrightarrow}\frac{1}{2}\int_{\text{BZ}}\frac{d^3k}{(2\pi)^3}F_k(\beta_c) \nonumber\\
=&&\ \ \frac{1}{2}\int_{-6t_f}^{6t_f}d\varepsilon N_{\text{3D}}(\varepsilon)\frac{\tanh\left(\frac{\beta_c\varepsilon}{2}\right)}{\varepsilon}
=\int_0^3 d\gamma N_{\text{3D}}(2t_f\gamma)\frac{\tanh(\beta_c t_f\gamma)}{\gamma} \nonumber\\
=&&\ \ \mathcal{N}_0 \bigg[\int_0^1 d\gamma \log(16\sqrt2)\frac{\tanh(\beta_c t_f\gamma)}{\gamma} + \int_1^3 d\gamma \frac{N_{\text{3D}}(2t_f\gamma)}{\mathcal{N}_0}\frac{\tanh(\beta_c t_f\gamma)}{\gamma}\bigg]   \nonumber \\
=&&\ \ \mathcal{N}_0\bigg[\frac{9}{2}\log2\int_0^3 d\gamma\frac{\tanh(\beta_c t_f\gamma)}{\gamma} + \int_1^3 d\gamma\left(\frac{N_{\text{3D}}(2t_f\gamma)}{\mathcal{N}_0}-\frac{9}{2}\log2\right)\underbrace{\frac{\tanh(\beta_c t_f\gamma)}{\gamma}}_{\approx\frac{1}{\gamma}}\bigg]   \nonumber \\
\stackrel{u=t_f\beta_c\gamma}{=}&&\ \ \mathcal{N}_0\bigg[\frac{9}{2}\log2\int_0^{3\beta_ct_f}du\frac{\tanh u}{u} + \underbrace{\int_1^3 d\gamma \left(\frac{N_{\text{3D}}(2t_f\gamma)}{\mathcal{N}_0}-\frac{9}{2}\log2\right)\frac{1}{\gamma}}_{=\kappa_1=-1.90} \bigg]   \nonumber \\
\approx&&\ \ \mathcal{N}_0\bigg[\frac{9}{2}\log2\log\left(\frac{12e^C}{\pi}\beta_ct_f\right)+\kappa_1\bigg],
\label{eq:phi1}
\end{eqnarray}
\end{widetext}
with $C\approx0.577$ the Euler-Mascheroni constant. Notice the low-$T$ approximation we did in the last step, as well as the way the non-universal constant $\kappa_1$ arises. We use the same method to calculate the function $\varphi_\gamma$, since $\gamma\sim\varepsilon$.

To evaluate the functions $\varphi_\eta$ and $\varphi_p$, we use a trigonometric trick, since the symmetry factors $\eta^2$ and $\sin^2 k_x$ cannot be directly written as a function of $\gamma$. Define a measure $d\mu(k)=\frac{d^3k}{2(2\pi)^3}F_k$ to integrate over the first Brillouin zone. Now, observe that for any fixed $i$ and any function $f = f(k_i)$ we have $\int d\mu(k)f(k_i) = \int d\mu(k)f(k_j)$ for any $i,j\in\{x,y,z\}$, since $F_k = F(\gamma_k)$, $\gamma_k = 2\sum_i\cos k_i$, and $k_i\in[-\pi,\pi]$. Then, we have
\begin{eqnarray}
 3\varphi_{\eta}+\varphi_{\gamma} =&&\ \ \int_{\text{BZ}}d\mu(k)3\eta_k^2+\gamma_k^2 \nonumber\\
 =&&\ \ \int_{\text{BZ}}d\mu(k) 4\bigg(4\left(\cos^2 k_x+\cos^2 k_y \right)+\cos^2 k_z \nonumber\\
  -&&\ \ \underline{4\cos k_x\cos k_y} + \underline{2\left(\cos k_x\cos k_z+\cos k_y\cos k_z\right)}\bigg) \nonumber\\
 \stackrel{k_i\leftrightarrow k_j}{=}&&\ \ \int_{\text{BZ}}d\mu(k) 36\cos^2k_x = 36(\varphi_1-\varphi_p),
\end{eqnarray}
where the underlined terms cancel each other out due to the aforementioned symmetry property. Hence, we see that $\varphi_p$ and $\varphi_\eta$ can be related using $\varphi_1$ and $\varphi_\gamma$ and one only needs to calculate $\varphi_p$. This is done along the same lines as Eq.~\ref{eq:phi1}, to arrive at Eqs.~(\ref{eq:UCS_phis0})-~(\ref{eq:UCS_phis}). For more details, see Ref.~\onlinecite{thesis}.

\end{appendix}


\begin{thebibliography}{99}

\bibitem{Greiner_Bloch}  M. Greiner, O. Mandel, T. Esslinger, T. E. H\"ansch, and I. Bloch, Nature (London) {\bf 39}, 415 (2002).

\bibitem{exp1} T. Best, S. Will, U. Schneider, L. Hackerm\"uller, D. van Oosten, I. Bloch, and D.-S.
L\"uhmann, Phys. Rev. Lett. {\bf 102}, 030408 (2009).

\bibitem{PhysRevLett.96.180403} S. Ospelkaus, C. Ospelkaus, O. Wille, M. Succo, P. Ernst, K. Sengstock, and K. Bongs, Phys. Rev. Lett. {\bf 96}, 180403 (2006).

\bibitem{PhysRevLett.101.050402} D.-S. L\"uhmann, K. Bongs, K. Sengstock, and D. Pfannkuche, Phys. Rev. Lett. {\bf 101}, 050402 (2008).

\bibitem{PhysRevB.80.054511} S. Tewari, R. M. Lutchyn, and S. Das Sarma, Phys. Rev. B {\bf 80}, 054511 (2009).

\bibitem{fermipolaron} S. Will, T. Best, S. Braun, U. Schneider, and I. Bloch, Phys. Rev. Lett. {\bf 106}, 115305 (2011).

\bibitem{PhysRevLett.96.180402} K. G\"unter, T. St\"oferle, H. Moritz, M. K\"ohl, and T. Esslinger, Phys. Rev. Lett. {\bf 96}, 180402 (2006).

\bibitem{PhysRevLett.100.140409} M. Cramer, S. Ospelkaus, C. Ospelkaus, K. Bongs, K. Sengstock, and J. Eisert, Phys. Rev. Lett. {\bf 100}, 140409 (2008).

\bibitem{PhysRevA.77.023608} L. Pollet, C. Kollath, U. Schollw\"ock, and M. Troyer, Phys. Rev. A {\bf 77}, 023608 (2008).

\bibitem{mering} A. Mering and M. Fleischhauer~, Phys. Rev. A {\bf 77}, 023601 (2008).

\bibitem{viverit2} L. Viverit and S. Giorgini, Phys. Rev. A {\bf 66}, 063604 (2002).

\bibitem{pieri1} E. Fratini, and P. Pieri Phys. Rev. A {\bf 88}, 013627 (2013); Phys. Rev. A {\bf 85}, 063618 (2012); Phys. Rev. A {\bf 81}, 051605(R) (2010).

\bibitem{pieri2} G. Bertaina, E. Fratini, S. Giorgini, and P. Pieri Phys. Rev. Lett. {\bf 110}, 115303 (2013).


\bibitem{ludwig} D. Ludwig, S. Floerchinger, S. Moroz, and C. Wetterich, Phys. Rev. A {\bf 84}, 033629 (2011).

\bibitem{Pollet2005} L. Pollet, M. Troyer, K. Van Houcke, and S. M. A. Rombouts, Phys. Rev. Lett. {\bf 96}, 190402 (2006).

\bibitem{Barillier2008} X. Barillier-Pertuisel, S. Pittel, L. Pollet, and P. Schuck, Phys. Rev. A {\bf 77}, 012115 (2008).

\bibitem{Hebert2007} F. H{\'e}bert, F. Haudin, L. Pollet, and G. G. Batrouni, Phys. Rev. A {\bf 76}, 043619 (2007).

\bibitem{Imambekov} A. Imambekov and E. Demler, Phys. Rev. A {\bf 73}, 021602R (2006).

\bibitem{Buchler2003} H. P. B{\"u}chler and G. Blatter, Phys. Rev. Lett. {\bf 91}, 130404 (2003).

\bibitem{Buchler2004} H. P. B{\"u}chler and G. Blatter, Phys. Rev. A {\bf69}, 063603 (2004).

\bibitem{bijlsma} M. J. Bijlsma, B. A. Heringa, and H. T. C. Stoof, Phys. Rev. A {\bf 61}, 053601 (2000).

\bibitem{heiselberg} H. Heiselberg, C. J. Pethick, H. Smith, and L. Viverit, Phys. Rev. Lett. {\bf 85}, 2418 (2000).

\bibitem{short_micnas} R. Micnas, J. Ranninger, S. Robaszkiewicz, and S. Tabor, Phys. Rev. B {\bf 37}, 9410 (1988).

\bibitem{footnote1} Notice that the phases of the BFM will be different for $U_{bf}<0$~\cite{ludwig}. At half-filling, however, the phase diagram of the BFM is symmetric w.r.t.~the hypersurface $U_{bf} = 0$. A similar situation occurs in the pure Fermi-Hubbard model, where due to the Lieb-Mattis transformation the phase diagram is symmetric w.r.t.~the line $U_{ff} = 0$ at half-filling.

\bibitem{stoof} H. T. Stoof,  D. Dickerscheid and K. Gubbels, Ultracold Quantum Fields, Springer (2009).

\bibitem{zhang} C.N. Yang and S.C. Zhang, Modern Physics Letters B {\bf 4}, 759-766 (1990).

\bibitem{esslinger} T. Esslinger, Annu. Rev. Condens. Matter Phys. {\bf 1}, p.129-152 (2010).

\bibitem{tamaki} H. Tamaki, Y. Ohashi, and K. Miyake, Phys. Rev. A {\bf 77}, 063616 (2008).

\bibitem{micnas} R. Micnas, J. Ranninger, and S. Robaszkiewicz, Rev. Mod. Phys. {\bf 62}, p.113-173 (1990).

\bibitem{ketterle} W. Ketterle and M. Zwierlein, Rivista del Nuovo Cimento {\bf 31}, p.247-422 (2008).

\bibitem{anders} P. Anders, P. Werner, M. Troyer, M. Sigrist, and L. Pollet, Phys. Rev. Lett. {\bf 109}, 206401 (2012).

\bibitem{footnote2} Notice that the presence of \emph{diagonal} long-range order in a \emph{superfluid} bosonic system manifests itself in a macroscopic occupation of the $\bf{k}=(\pi,\pi,\pi)$ mode, c.f.~App.~\ref{App1}. Therefore, to detect it one can use either one of the order parameters $\sum_i (-1)^i n_i$ and $b_{k=\pi}$. 

\bibitem{anders3} P. Anders, E. Gull, L. Pollet, M. Troyer, and P. Werner, New J. Phys. {\bf 13}, 075013 (2011).

\bibitem{iskin} M. Iskin, Phys. Rev. A {\bf 88}, 053606 (2013).

\bibitem{ostlund} A.B. Eriksson, T. Einarsson, and S. Ostlund, Phys. Rev. B {\bf 52}, 3662 (1995).

\bibitem{ostlund2} S. Ostlund, Phys. Rev. Lett. {\bf 69}, 1695 (1992).

\bibitem{thesis} M. Bukov, `Bose-Fermi Mixtures: A Mean-Field Study', Master Thesis, LMU M\"unchen, (2013).

\bibitem{bradlyn} B. Bradlyn, F. E. A. dos Santos, and A. Pelster, Phys. Rev. A {\bf 79}, 013615 (2009).

\bibitem{dossantos} F. E. A. dos Santos and A. Pelster, Phys. Rev. A {\bf 79}, 013614 (2009).

\bibitem{wang} D.-W. Wang, M. D. Lukin, and E. Demler, Phys. Rev. A {\bf 72}, 051604 (2005).

\bibitem{raghu} S. Raghu, S. A. Kivelson, and D. J. Scalapino, Phys. Rev. B {\bf 81}, 224505 (2010).

\bibitem{annett} J.F. Annett and J.P. Wallington, arXiv cond-mat/9807220, (1998).

\bibitem{efremov} D.V. Efremov and and L. Viverit, Phys. Rev. B {\bf 65}, 134519 (2002).

\bibitem{mathey} L. Mathey, S.-W. Tsai, and A. H. Castro Neto, Phys. Rev. Lett. {\bf 97}, 030601 (2006).

\bibitem{deforcrand} O. Akerlund, P. de Forcrand, A. Georges, and P. Werner, arXiv:1305.7136 (2013).

\bibitem{VCA} M. Potthoff, M. Aichhorn, and C. Dahnken, Phys. Rev. Lett. {\bf 91}, 206402 (2003).

\end{thebibliography}
\end{document}